\documentclass[prd,twocolumn,superscriptaddress,nofootinbib,showpacs]{revtex4}

\usepackage{graphicx,amsmath,amsfonts,amssymb,revsymb,dcolumn,epsfig,bm}

\def\be{\begin{equation}}
\def\ee{\end{equation}}
\def\bea{\begin{eqnarray}}
\def\eea{\end{eqnarray}}

\begin{document}
\title{Circles-in-the-sky searches and observable cosmic topology in a flat
Universe}

\author{B. Mota}\email{wronski@gmail.com}
\affiliation{Universidade Federal do Rio de Janeiro,
NACO - CCS - Av. Brigadeiro Trompowski s/n°\\
21941-590 Rio de Janeiro -- RJ, Brazil}

\author{M.J. Rebou\c{c}as}\email{reboucas@cbpf.br}
\affiliation{Centro Brasileiro de Pesquisas F\'{\i}sicas,
Rua Dr.\ Xavier Sigaud 150 \\
22290-180 Rio de Janeiro -- RJ, Brazil}

\author{R. Tavakol}\email{r.tavakol@qmul.ac.uk}
\affiliation{Astronomy Unit, School of Mathematical Sciences,
Queen Mary University of London\\ Mile End Road, London, E1 4NS, UK}

\date{\today}

\begin{abstract}
In a Universe with a detectable nontrivial spatial topology the
last scattering surface contains pairs of matching circles
with the same distribution of temperature fluctuations --- the
so-called circles-in-the-sky. Searches for nearly
antipodal circles-in-the-sky in maps of cosmic microwave background
radiation have so far been unsuccessful.
This negative outcome along with recent theoretical results
concerning the detectability of nearly flat compact topologies
is sufficient to exclude a detectable nontrivial topology for most
observers in \textit{very nearly flat} positively and negatively
curved Universes, whose total matter-energy density satisfies
$0<\mid\Omega_{\text{tot}}-1\mid \lesssim10^{-5}$.
Here we investigate the consequences of these searches for
observable nontrivial topologies if the Universe turns out
to be \textit{exactly flat} ($\Omega_{\text{tot}}=1$).
We demonstrate that in this case the conclusions
deduced from such searches can be radically different.
We show that, although there is no characteristic topological
scale in the flat manifolds, for all multiply-connected orientable
flat manifolds it is possible to directly study the action of the
holonomies in order to obtain a general upper bound on the angle that
characterizes the deviation from antipodicity of pairs of matching
circles associated with the shortest closed geodesic. This bound is valid
for all observers and all possible values of the compactification length parameters.
We also show that in a flat Universe there are observers for whom
the circles-in-the-sky searches already undertaken are insufficient
to exclude the possibility of a detectable nontrivial spatial topology.
It is remarkable how such small variations in the spatial curvature
of the Universe, which are effectively indistinguishable geometrically,
can have such a drastic effect on the detectability of cosmic topology.
Another important outcome of our results is that they offer a
framework with which to make statistical inferences from future circles-in-the-sky
searches on whether the Universe is exactly flat.
\end{abstract}


\maketitle

\section{Introduction}
Some fundamental open questions concerning the nature of our Universe are 
whether the Universe is spatially finite and what its shape and size are 
(see, e.g., the reviews~\cite{CosmTopReviews}).
An important point regarding these questions
is that the spatial geometry constrains but does
not determine the topology of the spatial sections $M$.
As a result, general relativity, as well as any local metrical theory of
gravitation, cannot determine the topology of the Universe, which
can in principle be found through observations.\footnote{In this work,
in line with the usage in the literature,
by topology of the Universe we mean the topology of its three
dimensional spatial sections $M$.}
A promising observational approach in the search for
possible evidence of a nontrivial cosmic topology is based on
searches for specific pattern repetitions in cosmic microwave
background (CMB) temperature fluctuations --- the so-called
circles-in-the-sky~\cite{CSS1998,Cornish-etal03,Key-et-al07,Aurich-etal2006}
(see also the related Refs.~\cite{RelatedCirc}).

The CMB data have become available through many experiments including the
ongoing Wilkinson Microwave Anisotropy Probe (WMAP)~\cite{wmap}.
The forthcoming data from CMB mission Planck~\cite{Planck-Collab},
which will be available in the near future,  will combine all-sky high
angular resolution and sensitivity with a wide frequency coverage,
and will certainly be a powerful data set to be used in the search
for a possible nontrivial topology of the Universe.
These accumulation of high precision CMB data
has at the same time provided strong support for the inflationary
scenario, and the near flatness of the Universe.

A fundamental assumption in standard relativistic
cosmological modelling is that the spacetime on large scales
is well described by a $4$--manifold $\mathcal{M} = \mathbb{R}\times M$
endowed with the spatially homogeneous and isotropic
Friedmann-Lema\^{\i}tre-Robertson-Walker (FLRW) metric
\be
\label{FLRW1}
ds^2 = -c^2dt^2 + a^2 (t) \left [ d \chi^2 + f^2(\chi) (d\theta^2 +
\sin^2 \theta  d\phi^2) \right ]\; ,
\ee
where $t$ is the cosmic time, $a(t)$ is the scale factor
and $f(\chi) = (\chi, \sin\chi,\sinh\chi)$ depending on the
sign of the constant spatial curvature $k=(0, 1, -1)$.
Furthermore, in the standard approach to modelling
the Universe the spatial sections $M$ are often assumed
to be the simply-connected $3$--manifolds:
Euclidean $\mathbb{E}^3$, spherical $\mathbb{S}^3$, or
hyperbolic space $\mathbb{H}^3$. These choices are, however, not unique,
and depending on the sign of the spatial curvature, the $3$-space $M$ can
be one of possible topologically-distinct $3$--manifolds, which
are quotient spaces of the corresponding simply-connected constant
curvature covering manifolds $\mathbb{E}^3$, $\mathbb{S}^3$ and
$\mathbb{H}^3$, by the group of isometries (the so-called holonomy group)
$\Gamma$ that define the set of closed  geodesics in each case.
The presence of closed geodesics in these quotient manifolds leads
to the existence of multiple images or pattern repetitions of
radiating sources. By observing these images or pattern repetitions
one can in principle directly obtain the elements of the holonomy
group and hence deduce the topology of the Universe.
Currently, the most promising method for searching for an observable non-trivial
topology is based on the existence of pattern repetitions in the CMB
anisotropies on the last scattering surface (LSS). In a Universe
with a detectable non-trivial topology~\cite{TopDetec} the LSS
intersects some of its images along the so-called circles-in-the-sky,
which are pairs of matching circles with the same distribution of
temperature fluctuations, identified by $\Gamma$, i.e.  with the
same distribution of temperature fluctuations (up to a phase).
Recent searches restricted to antipodal (back to back) or nearly
antipodal circles have been undertaken without
success~\cite{Cornish-etal03,Key-et-al07}.
Parallel to this it was proven in Ref.~\cite{Mota-etal} that in the case
of very nearly, but not exactly, flat manifolds (i.e., those compatible
with typical inflationary models), a generic compact manifold is
`locally' well approximated by either a slab space 
($\mathbb{R}^{2} \times \mathbb{S}^{1}$) or chimney space 
($\mathbb{R} \times \mathbb{T}^{2}$) manifold,
irrespective of its global topology.
This in turn allows an upper bound  to be placed on the angle
corresponding to the deviation from antipodicity in the inflationary
limit, which turns out to be less than $10$ degrees for the majority of
observers~\cite{Mota-etal-08}.
This result, coupled with the aforementioned searches,
which covered deviations from antipodicity of up to $10$ degrees,
would in principle be sufficient to exclude
a detectable manifold with non-trivial topology for
the overwhelming majority of observers in a very nearly flat
Universe.%
\footnote{Throughout this paper we assume that these searches would
reliably detect any pairs of matching circles present in the CMB data
as long as their parameters fall within the scope of the search (see also
Ref.~\cite{Aurich-et-al2008}).}

An important remaining question is how the above results
would be modified if the Universe turns out to be {\em exactly flat}.
Apart from being allowed by the observations, this limiting case
($\Omega_{\text{tot}}=1$)  is also significant, since it is compatible
with some inflationary scenarios. Indeed, it has been argued that flat
compact manifold should be regarded as typical for generic inflationary
scenarios (either along with hyperbolic compact manifolds~\cite{Linde}
or exclusively~\cite{McInnes}).

Here, to answer this question we
investigate what constraints the above mentioned existing searches
would impose on the detectability of topology in a flat Universe
in which the spatial section $M$ is an  Euclidean
space endowed with any orientable nontrivial topology.

In this paper, by considering all possible orientable multiply
connected flat $3$--manifolds, we show that the deviation from
antipodicity of the pair of circles-in-the-sky
can be, for some observers, larger than those expected
in the case of nearly flat Universes.
This therefore implies that,
for all globally inhomogeneous flat manifolds
with a nontrivial topology, there remains a substantial fraction
of observers for which the searches for the circles-in-the-sky
so far undertaken are not enough to exclude the possibility
of a detectable non-trivial flat topology.

The structure of the paper is as follows. In Section~\ref{Sec2} we
present a detailed study of the circles-in-the-sky in all multiply
connected flat orientable $3$--manifolds with a detectable topology,
and derive bounds on deviations from antipodicity of the circles in
each of these manifolds.
In Section~\ref{Sec3} we conclude with a brief discussion of the
significance of our results for possibility of detectable non-trivial
topology for the Universe in view of the present and future
CMB observations.

\section{Circles-in-the-sky in orientable flat 3-manifolds} \label{Sec2}

We wish to study the observable signatures of nontrivial topologies
of flat $3$--dimensional orientable manifolds. We begin by recalling that these
flat quotient manifolds are not \emph{rigid},
in the sense that topologically equivalent flat quotient manifolds,
defined by a given holonomy group $\Gamma$, can have different sizes.
Thus, there is no invariant characteristic topological scale with which one
can establish a detectability criterion (as given in Ref.~\cite{TopDetec},
for example).
Nevertheless, we know that the action of each pair of elements ($\gamma\,$,
$\gamma^{-1}$) of the group $\Gamma$ may generate one pair of
matching circles in the CMB maps when the LSS 
intersects its images under the action of  $\gamma\,$ and $\gamma^{-1}$.
Using this basic fact, it is possible to directly study the holonomies
in each orientable flat manifold with nontrivial topology in order to
obtain the maximum deviation from antipodicity of the pairs of matching
circles associated with the \emph{shortest closed geodesic} which contains
the observer's position in that manifold.%
\footnote{The  `end-points' of these geodesics, i.e. the observer's
position $\mathbf{p}$ and its nearest image $\gamma \mathbf{p}$,  are
the centers of the closest  neighboring images of the sphere of last
scattering (LSS).}
This specific pair of circles is important because if any other pair of
matching circles is detectable, then the pair associated with the shortest
closed geodesic will necessarily  also be detectable.%
\footnote{Of course, we are assuming in this article that all circles
that arise from the intersection of the LSS with any of its topological
copies are in practice statistically extractable from CMB maps.}
Conversely, if this pair from the closest images of the LSS is not
detectable, then we can be sure that no other pair of matching circles
will be detectable either.

We recall that in addition to the simply connected flat
Euclidean space $\mathbb{E}^{3}$, there are 17 multiply-connected
$3$--dimensional flat manifolds which are quotient spaces
of the form $\mathbb{E}^{3}/{\Gamma}$, where $\mathbb{E}^{3}$ is
the covering space and $\Gamma$ is the discrete and fixed point
free group of holonomies.
Only nine out of these quotient manifolds are orientable. These
consist of the six compact manifolds, namely $E_1$ ($3$--torus), $E_2$
(half turn space), $E_3$ (one quarter turn space), $E_4$ (one third turn space),
$E_5$ (one sixth turn space), $E_6$ (Hantzsche-Wendt space),
plus three non-compact ones: the chimney space $E_{11}$,
the chimney space with half turn $E_{12}$ and the slab spade $E_{16}$
(for details on the names and a description of the fundamental domain
of these manifolds see, e.g., Refs.~\cite{Adams-Shapiro01,Cipra02,Riazuelo-et-el03}).

The orientable manifolds $E_1$, $E_{11}$ and $E_{16}$ are globally
homogeneous and hence would only produce antipodal pairs of circles-in-the-sky.
The question then is what type of circles-in-the-sky would the
other remaining $6$ manifolds produce, and specifically what would be
the deviations from antipodicity for the pairs of  matching circles in each
of these manifolds,  and in particular for the circles associated with the
shortest closed geodesics.

Any holonomy $\gamma$ of an orientable Euclidean $3$--space can always
be expressed as a so-called screw motion (in the covering space),
consisting of a combination of a rotation $R(\alpha,\mathbf{\widehat{u}})$ by
an angle $\alpha$ around an axis of rotation $\mathbf{\widehat{u}}$,
plus a translation along the vector ${\bf L} = L \mathbf{\widehat{v}}$.
The action%
\footnote{The choice of axes to describe a screw motion is not unique,
but one can always find a rotation axis parallel to the direction
of translation.}
of $\gamma$ on any point ${\bf p}$ in the covering manifold is then
given by ${\bf p} \rightarrow R_{}^{}\,\mathbf{p} + \mathbf{L}$.
When there is no rotation part in the screw motion, i.e., when
$\alpha=0$, the holonomy is a translation, and its action
is exactly the same at every point; in particular the length
of the closed geodesic associated with $\gamma$,
$\ell_{\gamma} = \,\mid\gamma\,{\bf p}-{\bf p}\mid\, = L$, is the
same everywhere.
For a general screw motion, with $\alpha \neq 0$,  $\ell_{\gamma}$
depends on $\mathbf{p}$,  and in particular on the distance
between $\mathbf{p}$ and the axis of rotation.

The two matching circles associated with the holonomy
$\gamma = \left( R(\alpha,\mathbf{\widehat{v}}), {\bf L} \right)$ are
produced by the intersections of the sphere of last scattering with
its images under the isometries $\gamma$ and $\gamma^{-1}$.
The deviation from antipodicity, $\theta$, for any holonomy $\gamma$,
is given by
\begin{equation}
\label{theta}
\cos \theta =  - \,\frac{(\gamma {\bf p} - {\bf p}) \cdot (\gamma^{-1} {\bf p}
- {\bf p})}{\mid\gamma {\bf p} - {\bf p}\mid \,\, \mid\gamma^{-1} {\bf p}
- {\bf p}\mid}\,,
\end{equation}
where a dot denotes the usual scalar product in $\mathbb{E}^3$.

Let us first calculate $\theta$ for the manifolds $E_i,\,\,i=1, \cdots,5\,$,
which are quotient 3--manifolds that have as generators of the holonomy
group $\Gamma$ two translations plus a screw motion. In these manifolds,
the axis of rotation and the direction of the translation of the screw motion
are parallel. This common direction is perpendicular to the directions of
the translations associated with the other two remaining generators.
The rotation angle in these manifolds takes the form $\alpha=2\pi/n$,
where the so-called screw motion parameter $n$ is $1$, $2$, $4$,
$3$ and $6$ respectively. Of course, any translation will correspond
to the case $n=1$.

In order to obtain the value of $\theta_{\text{max}}$, the maximum deviation
from antipodicity between the pair of matching circles associated with the
shortest closed geodesic which contains the observer's position, we should in
principle compare the lengths of the closed geodesics generated by all elements
of $\Gamma$ at every point in the manifold. We note, however, that as the scale
of the translational generators is not fixed, the length of the smallest geodesic
corresponding to pure translations can be arbitrary.

Now if a pure translation, such as one of the translational generators with
axis perpendicular to the screw motion, happens to produce the shortest geodesic,
then its resulting circles-in-the-sky are back to back (zero deviation
from antipodicity). If, on the other hand, pure translations do
not produce the shortest geodesic, then we need to look at the remaining
holonomies which consist of the screw motion generator $\gamma_{SM}^{}$,
and its higher powers, $\gamma^m_{SM}$, for integer $m$. Since we want to obtain
an upper bound on $\theta$ applicable to all choices of topological
length parameter,
then clearly the configurations where this angle assumes its maximum value
are those in which the length of the closed geodesics associated with the
translational generators are  long enough so that the screw motion generator
and its powers always generate the shortest geodesics.
To obtain our upper bound we thus need to only consider this latter case.

Consider an arbitrary observer at a point ${\bf p}$, whose distance from
the axis of rotation is $r$. Without loss of generality, we can choose
coordinates $(\mathbf{x},\mathbf{y},\mathbf{z})$ such that the rotation
axis lies along the direction of $\mathbf{\widehat{x}}$, and the coordinates
of observer's position  are given by  ${\bf p} = (0,0,r)$. In this coordinate
system one has $\gamma_{SM}^{} {\bf p} = (L, r\sin \alpha, r \cos \alpha)$
and $\gamma_{SM}^{-1} {\bf p} = (-L, -r \sin \alpha, r \cos \alpha)$.
Then the lengths of the closed geodesics associated with
screw motion holonomies $\gamma_{SM}^m$ are given by
\begin{equation}
\ell_{\gamma_{SM}^m} \equiv
\left| {\gamma_{SM}^m} {\bf p} - {\bf p} \right|
= \sqrt{(mL)^2 + 2\,r^2\left( 1- \cos m\alpha \right)} \,\,.
\label{gx2}
\end{equation}

Now to decide which of these holonomies produces the smallest geodesic
for each point $\mathbf{p}$ in each of the considered five manifolds $E_i$,
we recall that the screw motion rotation angles are $\alpha=2\pi/n$ with
$n =1, 2, 4, 3, 6$. Note also that we need not concern
ourselves with the cases in which the translational generators produce
the shortest geodesic, since the maximum deviation from antipodicity for the
circles associated with the shortest geodesic will occur when the
(essentially arbitrary) parameters that define their lengths are large
enough so that the screw motion generates the shortest geodesic.

The question therefore now becomes: For which positive
integer $m$ is $\ell_{\gamma_{SM}^m}$ minimum? From Eq.~(\ref{gx2})
it is clear that if $m = s n + m'$, where $s$ is a
positive integer and $1 \leq m' \leq n$, then
$\ell_{\gamma_{SM}^{m'}} <  \ell_{\gamma_{SM}^{m}}$.
We can thus without loss of generality assume that
$1 \leq m \leq n$.

To proceed, we need to look inside the square root at the right hand side
of Eq.~(\ref{gx2}). It is the sum of two positive terms. The first, $(mL)^2$,
obviously increases with $m$ and is minimum for $m=1$. The second term,
$r^2\left( 1- \cos 2\pi m/n \right)$, assumes its minimum value when $m=n$
and its next-to-minimum value when $m=1$ (and $m=n-1$). Therefore, in all
cases $\ell_{\gamma_{SM}^m}$ is minimized for either $m=1$ or $m=n$, and
the length of the smallest geodesic is either $ \ell_{\gamma_{SM}^{}}$
or $\ell_{\gamma_{SM}^n}$. It is important to note that the latter length is
the same for all points of the manifold, since $\gamma_{SM}^n$ is a translation.

Now that we have narrowed down the holonomies we need to consider to only two,
$ \ell_{\gamma_{SM}^{}}$ and $\ell_{\gamma_{SM}^n}$, we only need to determine
for which points (as determined by the value of $r$) each
of the two holonomies under consideration generates the shortest geodesic.
To that end, we note that at the axis of rotation the
length associated with $\gamma_{SM}^{}$ is $\ell_{\gamma_{SM}^{}} = L$
(which is smaller than $\ell_{\gamma_{SM}^n} = n L$ for $n > 1$),
and this length increases monotonically as the distance $r$ from the axis
of the screw motion is increased, as can be seen from Eq.~(\ref{gx2}).
On the other hand, $\ell_{\gamma_{SM}^n}$ will of course remain fixed at $nL$.
Therefore, there is a sufficiently large $r$ such that the
equation
\begin{equation} \label{relation}
\ell_{\gamma_{SM}^{}} = \ell_{\gamma_{SM}^n} ,
\end{equation}
holds. This limiting case corresponds to the maximum value $r_{\text{max}}$
of the distance $r$  for which the shortest geodesic is generated by the screw motion.
Furthermore, by using the expression for the length given by Eq.~(\ref{gx2}) on both
sides of Eq.~(\ref{relation}) (for $m=1$ and $m=n$),
we obtain
\begin{equation}
n^2 = 1 + \frac{2\,r_{\text{max}}^2}{L^2} \left( 1-\cos \frac{2\pi}{n} \right) \,.
\label{rmax}
\end{equation}

On the other hand, to obtain $\theta_{\text{max}}$ we first substitute
the explicit expressions for  ${\bf p}$, $\gamma_{SM}^{} {\bf p}$
and $\gamma_{SM}^{-1} {\bf p}$ in the expression~(\ref{theta})
in order to relate the deviation from antipodicity
$\theta$ to the screw motion parameters $(\alpha= 2\pi/n, L)$
and the position of the observer (see also Ref.~\cite{German03}):
\begin{equation}
\cos \theta = 1 -
\frac{2\,r^2(1 - \cos \alpha)^2}{L^2+2\,r^2 (1 - \cos \alpha)} \,\,.
\label{theta-E1-5}
\end{equation}

This shows that the further away the observer is from the axis of rotation of
the screw motion, the greater is the deviation from antipodicity. As a result,
the maximum value of $r$ for which the shortest geodesic is generated by
$\gamma_{SM}$ instead of the translation $\gamma_{SM}^n$ corresponds to
the maximum value of $\theta$, $\theta_{\text{max}}$.

We can now obtain $\theta_{\text{max}}$ for the manifolds $E_i,\,\,i=1,...,5$ and
$E_{12}$  by substituting the value of $r_{\text{max}}$ given by Eq.~(\ref{rmax})
into Eq.~(\ref{theta-E1-5}) to obtain
\begin{equation}
\cos(\theta_{\text{max}})=1-\frac{n^2-1}{n^2} \left[1 - \cos \left(\frac{2\pi}{n}
     \right) \right]\,,\label{thetamax}
\end{equation}
which readily allows the calculation of the values of
$\theta_{\text{max}}$ for each of these manifolds.

We note that the results concerning the manifolds
$E_i,\,\,i=1, \cdots,5$ also apply to the chimney with half turn
$E_{12}$, as this manifold can be considered as a limiting case of $E_2$
(half turn space) when the length parameter of one of the translational
generators goes to infinity. Furthermore, the chiney  and slab spaces,
$E_{11}$ and $E_{16}$, are clearly the limiting cases of the
three-torus $E_{1}$ in which one or two of the length parameters
go to infinity.

\begin{table}[ht!]
\begin{tabular}{*4{c}}  
 \hline\hline
Symbol   & Manifold                     &    $n$   & \hspace{2mm} $\theta_{\text{max}}$  \\
\hline
$E_1$    &   three-torus                & 1,1,1  & \hspace{2mm} $0^{\circ}$    \\
$E_2$    & half turn space              & 1,1,2  & \hspace{2mm} $120^{\circ}$  \\
$E_3$    & quarter turn space           & 1,1,4  & \hspace{2mm} $86^{\circ}$   \\
$E_4$    & third turn space             & 1,1,3  & \hspace{2mm} $109^{\circ}$  \\
$E_5$    & sixth turn space             & 1,1,6  & \hspace{2mm} $59^{\circ}$   \\
$E_6$    & Hantzsche-Wendt space        & 2,2,2   & \hspace{2mm} $120^{\circ}$  \\
$E_{11}$ & chimney space                & 1,1  & \hspace{2mm} $0^{\circ}$    \\
$E_{12}$ & chimney space with half turn & 1,2 & \hspace{2mm}  $120^{\circ}$ \\
$E_{16}$ & slab space                   & 1  & \hspace{2mm} $0^{\circ}$    \\
\hline\hline
\end{tabular}
\caption{Multiply-connected flat orientable manifolds and the
maximum deviation from antipodicity of the circles-in-the-sky
for each manifold. The screw motion twist parameters $n$
for all the generators are also indicated.} \label{table:theta-max}
\end{table}

In Table~\ref{table:theta-max} we collect the values of $\theta_{\text{max}}$, the maximum
possible deviation from antipodicity of the circle pairs associated
with the shortest closed geodesic (i.e., the most readily detectable),
for all orientable flat three-manifolds with a nontrivial topology. We note that the
calculation of $\theta_{\text{max}}$ for $E_6$ (Hantzche-Wendt space)
is more involved and is presented in detail in the Appendix.
This Table shows that although the flat manifolds are not rigid, a general maximum,
applicable to all observers and choices of length parameters, can be obtained for
the values of the deviation from antipodicity that needs to be included in a
comprehensive search for a detectable cosmic topology by the circles in the sky method.
The relatively large numerical values in the table make it apparent that the circles
searches so far undertaken~\cite{Cornish-etal03,Key-et-al07} (see also Ref.~\cite{RelatedCirc})
do not rule out the possibility of a detectable nontrivial flat topology.

Another important outcome of the results of Table~\ref{table:theta-max}
is that they offer a theoretical framework to draw conclusions from
future circles-in-the-sky searches regarding the presence or absence of
spatial curvature in the Universe. Given that
the deviation from antipodicity for a very nearly flat
($0<\mid\Omega_{\text{tot}}-1\mid \lesssim10^{-5}$) multiply connected
Universe is less than $10^\circ$ for the majority of
observers~\cite{Mota-etal-08},
the detection of a pair of circles-in-the-sky with a higher value of $\theta$
(e.g., $\theta \simeq 60^\circ$) would imply  that
the Universe is very likely to be flat (with $\Omega_{\text{tot}}=1$),
in the sense that in
nearly flat manifolds such high values for $\theta$ could only occur
for a vanishingly small subset of observers.
Note that no such definite determination can conceivably be made by
purely geometrical methods.

The use of the patterns of image repetitions in cosmic topology to constrain
spatial curvature has been discussed at some length in the literature
(see, e.g., Refs.~\cite{CosmTopReviews} and ~\cite{TopDetec}).
What we propose above is a direct observational test that may be able to
determine the curvature sign based on a a partial detection of the cosmic topology.
Indeed, it should not be surprising that flat and very nearly flat manifolds present
such radically different pictures regarding the detectable circles-in-the-sky patterns.
Although such manifolds may be geometrically indistinguishable (geometry being a
local property), they remain quite distinct topologically, since the possible
holonomy classes for zero, positive and negative spatial curvature are completely
different from each other (this is in turn a reflection of the differences in the
global properties of the covering spaces these groups tesselate). So, even locally,
the detectable topology contains information about the global shape of the Universe.

\section{Final Remarks} \label{Sec3}

The existence of correlated pairs of circles in the CMB anisotropy maps,
the so-called circles-in-the-sky, is a generic prediction of a detectable
non-trivial cosmic topology, regardless of the background geometry.
Detecting such circles, and measuring their position, angular radii
and relative phase, would allow us to characterize, and possibly determine,
the topology of the spatial section of the Universe.

Searches for circles-in-the-sky whose centers are separated by more
than $170^\circ$ (nearly antipodal circles)
have been performed with negative results. Recent theoretical results
together with this negative outcome would be sufficient to
exclude a detectable nontrivial topology for most observers in
\textit{very nearly flat} Universe.

In this work, we have studied what happens if the Universe turns out to be
\textit{exactly flat} ($\Omega_{\text{tot}}=1$), rather than nearly flat.
Despite the fact that flat manifolds  have no fixed
topological scales, we have been able, by studying the action of holonomies,
to derive upper bounds on the deviations from antipodicity
of the pairs of matching circles-in-the-sky associated with the
shortest closed geodesics in all orientable manifolds with a nontrivial
flat topology. The key point to bear in mind is that, if the circle pair
associated with the shortest closed geodesic is not detectable, then no
other circle pair will be detectable; but conversely, the failure to detect
any other (possibly back-to-back, or nearly so) pair of matching circles does
not guarantee that the pair of circles corresponding to shortest geodesic
is not detectable either. Thus, it can only be said that a given search is
comprehensive if it is designed to be able to detect the shortest geodesic's
circle pair for all observers. Our bounds can be regarded as defining what
constitutes a comprehensive search, in the idealized case in which detection
is certain if a detectable non-trivial topology exists.

The derived bounds show that the
searches already undertaken are not sufficient to exclude the possibility
of a nontrivial flat topology  for the Universe. Our results
also demonstrate that a slight variation in the value of the spatial
curvature by making it exactly flat, can have striking consequences
for the analysis of the detectability of cosmic topology.

Finally, the theoretical results presented here would allow us to make inferences
regarding the spatial curvature sign of the universe, in the case
of the detection of some correlated pairs of circles in upcoming
circles-in-the-sky searches.

\appendix*
\section{}
The the calculation of $\theta_{\text{max}}$ for  $E_6$ (Hantzche-Wendt
space) manifold is somewhat more involved, since the generators of its
holonomy group are three half-turn
screw motions (the choice of generators is not unique, but always include
at least three screw motions), and one needs to be careful in considering
all the possible combinations of the generators to find which holonomies
generate the shortest closed geodesic at each point.

More specifically, the action of the $E_6$ generators  can be expressed
as  three screw motions, $\gamma_1$, $\gamma_2$ and $\gamma_3$, consisting
of rotations of $\pi$ around respectively the $z$, $x$ and $x$-axes,
followed by translations of, respectively, $a$ along the z-axis, $b$ along
the z-axis plus $c$ along the $y$-axis and $-b$ along the $z$-axis plus $-c$
along the $y$-axis respectively. Thus, the actions of the generators on a
point ${\bf p}$ on the manifold is given by
\bea
\gamma_1{\bf p} &=& R(\pi, \hat z){\bf p} + a \hat z \nonumber \\
\gamma_2{\bf p} &=& R(\pi, \hat x){\bf p} + b \hat x + c \hat y  \nonumber \\
\gamma_3{\bf p} &=& R(\pi, \hat x){\bf p} + -b \hat x - c \hat y.
\eea

The key  point to note is that, since $\gamma_1$, $\gamma_2$ and $\gamma_3$
are all half-turn screw motions around one of the coordinate axes, then
any combination of these holonomies and their inverses will also be either
a half-turn screw motion around one of the coordinate axis or a
translation (a "full turn" screw motion).%
\footnote{To see this, note that generally the combination of screw motions
is itself a screw motion, with an $R$ matrix that is the product of the
component $R$ matrixes. Since in this case the $R$ matrixes for the generators
are diagonal, with the diagonal elements being either $1$ or $-1$ such that $\det(R)=1$,
then the diagonal elements and determinant for the resulting $R$ matrix will be simply
the product of respectively the diagonal elements and determinants of the generators' $R$
matrixes. In other words, the resulting $R$ matrix will be diagonal, with elements being
either $1$ or $-1$, and $\det(R)=1$, i.e., $R$ will be either a rotation of $\pi$
around one of the coordinate axes or the identity matrix.}
In the case of a half-turn screw motion, the translation axis will not
in general be parallel to the axis of rotation. But it is well known
(see e.g. Ref.~\cite{Mota-etal})
that for a suitable change of axis any screw motion can be expressed as
a rotation  followed by a translation along the appropriately chosen axis.
Moreover, two equivalent screw motions will have parallel
axes and equal values for the rotation angle.

Now, let us assume that for an observer with position ${\bf p}$, and
for some specific values of the constants $a$, $b$ and $c$, the
holonomy ${\gamma_S}$ generates the shortest closed geodesic.
From the brief discussion above, it is clear that $\gamma_S$ is
either a translation, in which case the deviation from antipodicity
for the most readily  detectable pair of circles is $0$, or $\gamma_S$
is a half-turn screw motion. In the latter case $\gamma_S$ consists
of a rotation around some axis $\hat w$ (parallel to one of the coordinate
axes) plus a translation of $L$ along $\hat w$:
$\gamma_S{\bf p} = R(\pi, \hat w){\bf p} + L \hat w$.
The length of the associated closed geodesic is given by Eq.~(\ref{gx2})
(with $m=1$ and $\alpha=\pi$). Furthermore, the holonomy
$\gamma_S^2 {\bf p} = {\bf p} + 2 L \hat w$ is a translation
with length $2L$. Thus, by the same reasoning we applied to the
other manifolds, the maximum deviation from antipodicity for which
$\gamma_{S}$ generates a shorter closed geodesic than $\gamma_S^2$
occurs when $\ell_{\gamma_{S}^{}} = \ell_{\gamma_{S}^2}$.
Performing the same calculation we used to estimate $\theta_{\text{max}}$
for $E_2$, we obtain from Eq.~(\ref{thetamax}) (with $n=2$) that
$\theta_{\text{max}}^{E_6} \leq 120^{\circ}$.

Finally, all that remains is to prove that there exists at least one
combination of $a$, $b$, $c$ and {\bf p} such that the deviation
$\theta$ from antipodicity associated with the holonomy
generating the shortest geodesic is exactly $120^{\circ}$.
To show  that this is indeed the case, note that for
$b>2a$ and $c>2a$, either $\gamma_1$ or $\gamma_1^2$ will
generate the shortest closed geodesic at any position, and that {\bf p}
can be chosen so that $\ell_{\gamma_{1}^{}} = \ell_{\gamma_{1}^2}$,
which again from Eq.~(\ref{thetamax}) results in $\theta=120^{\circ}$.
Since $\theta_{\text{max}}$ is by definition the maximum value of $\theta$ for
all possible parameter and position combinations, this implies that
$\theta_{\text{max}}^{E_6} \geq 120^{\circ}$.

By combining both inequalities, we then obtain that
$\theta_{\text{max}}^{E_6} = 120^{\circ}$.

\begin{acknowledgments}
This work is supported by Conselho Nacional de Desenvolvimento
Cient\'{i}fico e Tecnol\'ogico (CNPq) -- Brasil, under grant No.\
472436/2007-4. M.J. Rebou\c{c}as and R. Tavakol thank, respectively,
CNPq and PCI-CBPF/MCT/CNPq for the grants under which this work was
carried out. B. Mota and M.J. Rebou\c{c}as also thank the Astronomy
Unit of QMUL for hospitality during a visit when part of this work
was carried out.
\end{acknowledgments}



\end{document}